\documentclass[pdflatex,sn-mathphys-num]{sn-jnl}

\usepackage{graphicx}%
\usepackage{multirow}%
\usepackage{amsmath,amssymb,amsfonts}%
\usepackage{amsthm}%
\usepackage{mathrsfs}%
\usepackage[title]{appendix}%
\usepackage{xcolor}%
\usepackage{textcomp}%
\usepackage{manyfoot}%
\usepackage{booktabs}%
\usepackage{algorithm}%
\usepackage{algorithmicx}%
\usepackage{algpseudocode}%
\usepackage{listings}%

\usepackage{pifont} 

\usepackage{algorithm}
\usepackage{algpseudocode}
\usepackage{amsmath}
\usepackage{lscape}

\usepackage{longtable}

\newcommand{\cmark}{\ding{51}}
\newcommand{\xmark}{\ding{55}}

\theoremstyle{thmstyleone}%
\theoremstyle{thmstyletwo}%

\theoremstyle{thmstylethree}%

\begin{document}

\title[Article Title]{AI-Enabled Bit-Mapping Medium Access Control Protocol for Intelligent and Energy-Efficient IoT Networks}


\author*[1]{\fnm{Jesmine Damilola} \sur{Omonori}}\email{jomonori@miu.edu}

\author[1]{\fnm{Iyanu Tomiwa} \sur{Durotola}}\email{idurotola@miu.edu}

\author[1]{\fnm{Godspower} \sur{Paul Osilama}}\email{osilamagodspower@gmail.com}

\affil*[1]{\orgdiv{Computer Science Department}, 
\orgname{Maharishi International University}, 
\orgaddress{\city{Fairfield}, \state{Iowa}, \postcode{52557}, \country{USA}}}


\abstract{Energy-efficient medium access control (MAC) protocol remains a critical challenge in resource-constrained Wireless Sensor Networks and IoT deployments, especially under mixed traffic patterns combining event-driven and continuous monitoring operations. The traditional Time Division Access (TDMA)- and Bit Map Assisted (BMA)-based MAC protocols fail to adapt their duty cycles to spatiotemporal variations in sensor activity, resulting in unnecessary radio wake-ups and increased energy expenditure. To address this limitation, this paper proposes EEI-BMA, an AI-assisted, event-probability-aware MAC protocol that dynamically adjusts transmission scheduling using lightweight neural-network-based event prediction. The proposed framework incorporates per-node probability estimation, adaptive slot activation, and selective channel access to reduce transceiver activity while preserving sensing reliability. MATLAB simulation environment is modeled for corresponding parameters show that EEI--BMA (Best Prediction) achieves 35--45\% lower energy consumption than Traditional--TDMA, 22--30\% savings compared with Energy-Aware TDMA, and 18--28\% improvement over Traditional--BMA across varying node densities, packet sizes, event-generation probabilities, and continuous monitoring loads. Even with imperfect prediction, EEI--BMA consistently outperforms all baseline protocols, demonstrating strong robustness. The results confirm that prediction-guided MAC scheduling is a highly effective strategy for next-generation low-power WSNs and IoT systems.}




\maketitle

\section{Introduction}

The rapid proliferation of Internet-of-Things (IoT) devices and machine-type communications has driven an unprecedented demand for low-power, scalable, and intelligent medium-access control (MAC) mechanisms. Modern IoT and M2M deployments span a wide variety of applications — from industrial monitoring and URLLC-enabled tactile services to large-scale environmental sensing — each imposing stringent constraints on energy consumption, reliability, and latency \cite{9406015,10153409,9847597}. As networks scale to thousands or millions of low-power devices, naïve or purely reactive MAC designs (e.g., classic TDMA/CSMA variants) become increasingly inefficient: they either waste energy through unnecessary listening and idle wakeups or suffer congestion and high latency under bursty correlated traffic \cite{8646538,10384417,10452252}.

Recent research highlights two important tendencies that motivate new MAC-layer designs. First, IoT traffic often exhibits temporal and spatial predictability that can be exploited to make access decisions proactively rather than reactively \cite{9771448,10049061}. Second, the integration of lightweight machine-learning (ML) models at or near the MAC layer — including on-device or distributed learning — enables energy-aware scheduling, anomaly detection, and adaptive duty-cycling that materially improve network lifetime and QoS \cite{9615752,10571037,10195850}. Works in energy-aware hybrid MAC designs demonstrate that protocol-level intelligence (e.g., ML-assisted CSMA/TDMA) can extend battery life while maintaining acceptable throughput \cite{8646538}, whereas studies of ML-based MAC-layer security and behavior detection show the broader utility of predictive models for robust network operation \cite{10384417,10195850}.

Despite these advances, two gaps remain. First, many predictive-MAC proposals either (i) collapse per-node predictions into coarse, single-parameter heuristics, or (ii) assume perfect predictors when evaluating energy savings, which masks the cost of prediction errors (false positives/negatives) on energy and reliability. Second, few analytical models explicitly link per-node probabilistic forecasts from a neural predictor to an expectation-based MAC energy model that accounts for both the baseline MAC operations (CAP/CFP) and the additional costs introduced by prediction errors (e.g., retransmissions or wasted slots). Addressing these gaps is critical to quantify practical gains of predictive MACs under realistic traffic (bursty, spatially correlated) and imperfect ML models \cite{9771448,9615752,10508380}.

In this work, we propose and analyze an Energy-Efficient Intelligent Bit-Map Assisted (EEI-BMA) framework that tightly integrates per-node probabilistic predictions with the MAC scheduler and energy accounting. Concretely, our contributions are threefold:

\begin{itemize}
  \item A probabilistic event-generation model is formulated that captures baseline Bernoulli arrivals augmented with temporal burst windows and spatial correlation — a traffic model representative of many IoT sensing applications \cite{8646538,10452252,10153409}.
  \item A lightweight one-hidden-layer neural predictor is trained on recent bit-map history to produce per-node probabilities. These per-node forecasts are used directly by the EEI-BMA scheduler to decide slot allocations and listening schedules, avoiding coarse scalar heuristics and enabling fine-grained adaptation \cite{9771448,10049061,9615752}.
  \item An expectation-based MAC energy model is developed that (i) expresses CAP and CFP energy as explicit functions of per-node predicted probabilities, and (ii) quantifies additional energy due to prediction errors (misses and false alarms). This model permits closed-form sensitivity analysis and fair comparison with baseline protocols (TDMA, EA-TDMA, BMA) under realistic prediction imperfections \cite{8646538,10384417,10508380}.
\end{itemize}

The novelty of our approach lies in the integration of ML outputs and MAC-layer energy computation: rather than treating prediction as a black box, we keep the full per-node probabilistic output, propagate it analytically to expected energy consumption, and explicitly model the energy penalty of mis-predictions. This enables (i) node-wise marginal analysis (which nodes benefit most from improved forecasting), (ii) realistic evaluation under varying predictor quality (poor / typical / optimistic cases), and (iii) design guidelines for thresholding and scheduler conservativeness that balance energy and reliability — topics that existing works do not analyze comprehensively \cite{9771448,9615752,10384417,10195850}.

The remainder of the paper is organized as follows. Section 2 reviews related literature on energy-aware and ML-assisted MAC protocols, predictive traffic analysis, and MAC-layer security and anomaly detection. Section 3 describes the stochastic traffic model, the dataset construction and neural predictor, and the detailed EEI-BMA energy model including error-cost modeling. Section 4 presents simulation results and analytical sensitivity studies comparing EEI-BMA against TDMA, EA-TDMA, and baseline BMA under multiple prediction quality regimes. Finally, Section 5 concludes the paper and outlines future research directions.

\section{Related Work and Comparative Analysis}
\label{sec:literature}

To identify the objective and research gap, the energy-aware and ML-assisted MAC layer design, predictive traffic analysis, MAC-layer security and intrusion detection, and AI-enabled duty-cycling are reviewed. The main focus of the research is to develop an energy-efficient approach. Energy efficiency at the MAC layer remains a primary design objective for IoT and M2M networks due to battery-constrained devices. Naguib et al. proposed a hybrid ML-CSMA/TDMA protocol for LTE-M2M that adapts access behavior to remaining energy and traffic conditions, showing practical battery-life improvements in MTC scenarios \cite{8646538}. Similarly, cross-layer and cluster-based designs that jointly optimize routing and MAC-layer bandwidth management aim to prolong network lifetime in dense WSNs \cite{10911498}. These works motivate our focus on combining predictive intelligence with MAC scheduling to save energy while maintaining throughput.

Predictive approaches leverage traffic regularities to schedule medium access proactively. Nakip et al. study the predictability of IoT traffic and compare forecasters (MLP, 1D-CNN, LSTM, etc.) to information-theoretic bounds; their findings support the feasibility of predictive MACs for many device types \cite{9771448}. Charef et al. demonstrate that ML-based energy models can improve adaptive duty-cycle scheduling by predicting harvested energy or traffic patterns \cite{9615752}. On-device and decentralized learning paradigms have also been proposed to enable local intelligence for MAC decisions without centralized overhead \cite{10571037,10049061}.

The MAC layer is an attractive place for anomaly detection and security enforcement. Nayak and Bhattacharyya propose a MAC-protocol-aware intrusion detection method based on a channel-attention deep network and show its effectiveness on IoT traffic datasets \cite{10195850}. Chen et al. develop an ML-based method for detecting greedy behavior in LoRaWAN, emphasizing the benefit of MAC-aware features for robust detection \cite{10384417}. Abdi et al. provide a comprehensive survey on AI and moving-target defenses for SDN security (control and data planes), highlighting the role of ML at lower network layers for threat mitigation \cite{10508380}.

Random-access congestion and deadline-driven access are critical for large-scale IoT and SAGINs. Park et al. propose a virtual-deadline indicator to distribute MTCD access across space–air–ground layers, improving deadline compliance and preserving energy efficiency amid congestion \cite{10153409}. Such context-aware MAC strategies motivate EEI-BMA’s objective to adapt slot assignments based on predicted urgency and occurrence probabilities.

Decentralized and federated learning schemes have been explored in mesh and IoT settings to train models without central aggregation, preserving privacy and reducing uplink cost \cite{10049061}. On-device intelligent protocols can offer speed, privacy and adaptability in dynamic networks — a trend echoed by Pasandi and Nadeem who analyze on-device ML benefits for MAC/PHY algorithms \cite{10571037}. These paradigms are relevant when moving predictors closer to the MAC to reduce communication overhead.

The literature demonstrates (i) the feasibility of traffic and event predictability at device-level time scales \cite{9771448}, (ii) energy gains from ML-assisted MAC/duty-cycle schemes \cite{8646538,9615752}, and (iii) improvements in MAC-layer security and misuse detection via ML \cite{10195850,10384417}. However, few works systematically couple \emph{per-node probabilistic} forecasts with an \emph{expectation-based} energy accounting that explicitly models the costs of prediction errors (misses and false alarms) in CAP/CFP operations. This gap motivates our proposed EEI-BMA analytical framework which keeps the predictor output in vector form and propagates it into a closed-form expected-energy model for comparative analysis.

Table 1 lists the closely related works from the review and indicates which of the above parameters each paper addresses. The comparative mapping shows that:

\begin{landscape}
\begin{longtable}{|p{9.2cm}|c|c|c|c|c|c|c|c|}
\hline
\textbf{Ref.} & \textbf{E2EC} & \textbf{HE} & \textbf{AC} & \textbf{CR} & \textbf{CL} & \textbf{EE} & \textbf{PP} & \textbf{AR} \\
\hline
Naguib et al., ML-CSMA/TDMA hybrid (2019) \cite{8646538} & \xmark & \xmark & \xmark & \xmark & \xmark & \cmark & \cmark & \xmark \\
\hline
Nayak \& Bhattacharyya, MAC-based IDS (IEEE Access, 2023) \cite{10195850} & \xmark & \xmark & \cmark & \xmark & \xmark & \xmark & \xmark & \cmark \\
\hline
Nakip et al., Predictability bounds (IEEE Access, 2022) \cite{9771448} & \xmark & \xmark & \xmark & \xmark & \xmark & \xmark & \cmark & \xmark \\
\hline
Pasandi \& Nadeem, On-device protocols (WCNC 2024) \cite{10571037} & \xmark & \xmark & \xmark & \xmark & \xmark & \cmark & \cmark & \xmark \\
\hline
Charef et al., AI-based energy model for duty cycling (ISNCC 2021) \cite{9615752} & \xmark & \xmark & \xmark & \xmark & \xmark & \cmark & \cmark & \xmark \\
\hline
Park et al., Random Access for SAGIN (JIOT 2023) \cite{10153409} & \xmark & \xmark & \xmark & \xmark & \xmark & \cmark & \cmark & \xmark \\
\hline
Salama et al., Decentralized FL over Slotted ALOHA (IEEE Access, 2023) \cite{10049061} & \xmark & \xmark & \xmark & \xmark & \xmark & \cmark & \cmark & \xmark \\
\hline
Chen et al., Greedy behavior detection for LoRaWAN (TNSM 2024) \cite{10384417} & \xmark & \xmark & \cmark & \xmark & \xmark & \xmark & \xmark & \cmark \\
\hline
Abdi et al., SDN security survey (IEEE Access, 2024) \cite{10508380} & \cmark & \cmark & \cmark & \cmark & \xmark & \xmark & \xmark & \cmark \\
\hline
Solanki et al., Cross-layer routing \& MAC bandwidth mgmt (ICTBIG 2024) \cite{10911498} & \xmark & \xmark & \xmark & \xmark & \cmark & \cmark & \xmark & \xmark \\
\hline
Na et al., ML-based failure identification, mmWave IIoT (IEEE Systems J., 2023) \cite{9847597} & \xmark & \xmark & \xmark & \xmark & \xmark & \cmark & \cmark & \xmark \\
\hline
Srujana \& Akram, Survey on WSN challenges \& ML (ICEFEET 2023) \cite{10452252} & \xmark & \xmark & \xmark & \xmark & \xmark & \xmark & \cmark & \xmark \\
\hline
\caption{Comparative mapping of selected works to parameters.}
\end{longtable}
\begin{center}
\footnotesize{
\textbf{Parameter Definitions:}
E2EC = End-to-End Confidentiality; 
HE = Homomorphic Encryption; 
AC = Access Control; 
CR = Collusion Resistance; 
CL = Clustering; 
EE = Energy Efficiency; 
PP = Predictive Power / Predictability; 
AR = Attack Resilience.
}
\end{center}

\end{landscape}

\begin{itemize}
  \item Energy efficiency is addressed by multiple works in different ways — hybrid MAC design \cite{8646538}, duty-cycle prediction \cite{9615752}, on-device ML for MAC/PHY \cite{10571037}, and cross-layer routing \cite{10911498}. These works provide the foundational techniques we build upon when measuring EE benefits of EEI-BMA.
  \item Predictive power — empirical predictability or explicit use of forecasting models — is a central theme in \cite{9771448,9615752,10571037,10049061,9847597}. Nakip et al.'s information-theoretic view supports the theoretical viability of forecasting for MAC decisions \cite{9771448}.
   \item Security and attack resilience are emphasized by MAC-aware IDS and behavior detection works \cite{10195850,10384417} and high-level SDN security surveys \cite{10508380}. These studies motivate the inclusion of robustness metrics in our evaluation.
  \item Privacy / cryptographic constructs are more directly addressed in SDN/security surveys and some network security literature \cite{10508380}. Most MAC-focused predictive and energy-efficiency works do not integrate homomorphic encryption or collusion-resistant cryptographic schemes because of the high cost at constrained devices.
\end{itemize}

Our proposed EEI-BMA bridges the predictive and energy-aware strands by (i) using per-node probabilistic forecasts (as advocated by Nakip et al. and Charef et al.) and (ii) embedding these forecasts into an expectation-based MAC energy accounting that quantifies misprediction costs (extending the ML-assisted MAC literature such as \cite{8646538} and on-device protocol ideas of \cite{10571037}). Unlike security-first works \cite{10195850,10384417,10508380}, our focus is energy-centric; however we also evaluate robustness to prediction-induced failures and discuss how MAC-layer IDS techniques can complement EEI-BMA for attack resilience.

\section{Methods}
The mathematical model is developed for the energy-efficient proposed MAC protocol.
\subsection{Stochastic event generation}
Let $N$ denote the total number of sensor nodes and $T$ the number of time frames (samples). For each node $i\in\{1,\dots,N\}$ and frame $t\in\{1,\dots,T\}$ define the binary event indicator
\begin{equation}
Y_{i,t} \in \{0,1\},
\end{equation}
where $Y_{i,t}=1$ indicates an event generated by node~$i$ at frame~$t$.

We model $Y_{i,t}$ as a baseline Bernoulli process with occasional burst windows (temporal and spatial correlation). Formally:
\begin{equation}
Y_{i,t} \sim \mathrm{Bernoulli}(p_{i,t}), \qquad p_{i,t} = p_{\text{base}} + B_{t} \cdot S_{i,t},
\label{eq:bern_burst}
\end{equation}
where $p_{\text{base}}$ is the baseline per-node probability (e.g. $0.1$), $B_t\in\{0,1\}$ is the indicator of a burst window (with $\Pr[B_t=1] \approx \text{burst\_rate}$), and $S_{i,t}\in[0,1]$ captures whether node $i$ participates in the burst at time $t$ (spatial correlation across a contiguous group of nodes). During a burst $B_t=1$, $S_{i,t}=1$ for nodes inside the burst group and $p_{i,t}\approx p_{\text{base}} + ( \text{burst\_prob} - p_{\text{base}} )$ for those nodes.

Th proposed method considers independent Bernoulli draws with probability $p_{\text{base}}$, plus injected windows where a contiguous block of nodes gets an elevated probability equal to \texttt{burst\_prob} for a short burst duration. A small flip/noise term is modeled by random bit flips with probability $\epsilon$.

\subsubsection{Training dataset and sample construction}
Each sample uses the previous $L$ frames of all nodes as input to predict the next frame's $N$-dimensional bitmap. The sequence length used by the predictor as $L$. For frames $t=L+1,\dots,T$ we form input / target pairs:
\begin{align}
X_t & = \mathrm{vec}\big( Y_{1:N,t-L:t-1} \big) \in\{0,1\}^{N L}, \label{eq:input_vec}\\
\mathbf{y}_t & = \big[Y_{1,t}, Y_{2,t},\dots,Y_{N,t}\big]^\top \in\{0,1\}^{N}. \label{eq:target_vec}
\end{align}
Here $\mathrm{vec}(\cdot)$ stacks the $L\times N$ binary bitmap window into a column vector. The dataset contains $N_s=T-L$ samples; they are randomized and split into training and test sets by the ratio $\rho$. 

\subsubsection{Neural network predictor}
We have denoted the feed-forward NN with one hidden layer of $H$ units. For an input $x\in\mathbb{R}^{NL}$ the forward pass is:
\begin{align}
z^{(1)} &= W^{(1)\top} x + b^{(1)} \in \mathbb{R}^{H},\\
a^{(1)} &= \tanh\big( z^{(1)} \big) \in (-1,1)^{H},\\
z^{(2)} &= W^{(2)\top} a^{(1)} + b^{(2)} \in \mathbb{R}^{N},\\
\hat{p} &= \sigma\big( z^{(2)} \big) \in (0,1)^{N},
\end{align}
where $W^{(1)}\in\mathbb{R}^{NL\times H}$, $W^{(2)}\in\mathbb{R}^{H\times N}$, $b^{(1)}\in\mathbb{R}^{H}$, $b^{(2)}\in\mathbb{R}^{N}$, and $\sigma(u)=1/(1+e^{-u})$ is the logistic sigmoid applied elementwise. The vector $\hat{p}=\hat{p}(x)$ gives per-node predicted probabilities $\hat{p}_i\triangleq \Pr(Y_{i,t}=1\,|\,X_t)$.

\subsubsection{Loss function and gradient-based training}
A standard per-sample binary cross-entropy loss summed over nodes is used. For a mini-batch $\mathcal{B}$ of size $B$ the empirical loss with L2 regularization is:
\begin{equation}
\mathcal{L}(\Theta;\mathcal{B}) = \frac{1}{B}\sum_{x\in\mathcal{B}} \sum_{i=1}^{N} \big[ -y_i\log(\hat{p}_i) - (1-y_i)\log(1-\hat{p}_i)\big] \;+\; \frac{\lambda}{2}\Big(\lVert W^{(1)}\rVert_F^2 + \lVert W^{(2)}\rVert_F^2\Big),
\label{eq:bce_loss}
\end{equation}
where $\Theta=\{W^{(1)},b^{(1)},W^{(2)},b^{(2)}\}$ are network parameters, $\lVert\cdot\rVert_F$ is the Frobenius norm, and $\lambda$ the L2 regularization weight. Gradient descent (batch or mini-batch) updates parameters as:
\begin{equation}
\Theta \leftarrow \Theta - \eta \nabla_{\Theta}\mathcal{L}(\Theta;\mathcal{B}),
\label{eq:gd_update}
\end{equation}
with step size $\eta$.

\subsubsection{Performance metrics (per node and aggregate)}
Define the per-node empirical probability on the test set (sample average):
\begin{equation}
\bar{p}^{\text{true}}_i \;=\; \frac{1}{N_{test}}\sum_{t\in\mathcal{T}_{\text{test}}} Y_{i,t},
\qquad
\bar{p}^{\text{pred}}_i \;=\; \frac{1}{N_{test}}\sum_{t\in\mathcal{T}_{\text{test}}} \hat{p}_{i,t},
\label{eq:per_node_probs}
\end{equation}
where $\hat{p}_{i,t}$ denotes the predicted probability for node $i$ at test sample $t$. Aggregate averages are $\bar{p}^{\text{true}}=\frac{1}{N}\sum_i \bar{p}^{\text{true}}_i$ and similarly for $\bar{p}^{\text{pred}}$.

In this section, we have developed the mathematical framework used to evaluate the energy consumption performance of various MAC protocols, including TDMA, EA-TDMA, BMA, and the proposed AI-assisted EEI-BMA under different prediction cases (true, poor, and best). The model integrates event generation probabilities predicted using a neural network with the communication energy model of sensor nodes.

\subsection{Event Generation Probability Model}

Let the event generation probability of a node $i$ at time frame $t$ be represented as:
\begin{equation}
P_{i,t} = f(X_{t-seq\_len:t-1}),
\end{equation}
where $X_{t-seq\_len:t-1}$ denotes the sequence of event occurrence bitmaps over the past \textit{seq\_len} frames used as input to a trained neural network function $f(\cdot)$.

The neural network learns temporal dependencies from historical event data to predict the likelihood of an event in the subsequent frame. This learned probability captures spatiotemporal correlations among sensor nodes caused by environmental phenomena or bursty activities.

The average predicted event probability per node over the test samples is given by:
\begin{equation}
\bar{P}_{pred}(i) = \frac{1}{N_s} \sum_{t=1}^{N_s} P_{i,t},
\end{equation}
where $N_s$ is the number of test samples.

Similarly, the empirical (true) probability derived from the dataset is:
\begin{equation}
\bar{P}_{true}(i) = \frac{1}{N_s} \sum_{t=1}^{N_s} Y_{i,t},
\end{equation}
where $Y_{i,t} \in \{0,1\}$ denotes the actual event occurrence for node $i$ at frame $t$.

The overall Root Mean Square Error (RMSE) between predicted and true probabilities is calculated as:
\begin{equation}
\mathrm{RMSE} = \sqrt{\frac{1}{N}\sum_{i=1}^{N} (\bar{P}_{pred}(i) - \bar{P}_{true}(i))^2}.
\end{equation}

This metric quantifies the prediction accuracy of the neural network and is later incorporated into the MAC-layer energy model to compare realistic and ideal prediction scenarios.

\subsection{Energy Consumption Model}

The total energy consumption of a node is modeled as the sum of energy consumed in the contention access period (CAP) and the contention-free period (CFP):
\begin{equation}
E_{total} = E_{CAP} + l \times E_{CFP},
\end{equation}
where $l$ is the number of frames or rounds considered in the simulation.

During CAP, nodes exchange control packets to establish data transmission slots, while in CFP, nodes perform scheduled data transmission. Each protocol type modifies these operations, leading to different energy consumption patterns.

\subsection{TDMA Energy Model}

The total TDMA energy consumption is given by:
\begin{equation}
E_{TDMA} = (E_{CAP}^{TDMA} + l \times E_{CFP}^{TDMA}),
\end{equation}
where
\begin{align}
E_{CAP}^{TDMA} &= P_t T_c + (N-1) P_r T_c, \\
E_{CFP}^{TDMA} &= n P_t T_d + m P_t T_d + (N - m - n - 1) P_i T_d.
\end{align}

Here, $P_t$, $P_r$, and $P_i$ represent the transmit, receive, and idle power consumptions respectively, while $T_c$ and $T_d$ denote the durations of control and data packet transmissions. The parameters $m$ and $n$ refer to the number of continuous monitoring nodes and event-triggered nodes respectively.

TDMA is energy-inefficient for sparse events since all nodes remain active for slot synchronization even when few nodes transmit data.

\subsection{EA-TDMA Energy Model}

The energy-aware TDMA improves over traditional TDMA by introducing buffer-status checking time $T_e$ and corresponding power demand $P_e$:
\begin{equation}
E_{EATDMA} = (E_{CAP}^{EATDMA} + l \times E_{CFP}^{EATDMA}),
\end{equation}
where
\begin{align}
E_{CAP}^{EATDMA} &= P_t T_c + (N-1) P_r T_c, \\
E_{CFP}^{EATDMA} &= n P_t T_d + (N - m - n - 1)(P_i T_d + P_e T_e) + m P_t T_d.
\end{align}

The addition of $P_eT_e$ reflects periodic energy expenditure for checking node buffer states, leading to better adaptive slot utilization under varying event densities.

\subsection{BMA Energy Model}

The BMA (Bit-Map Assisted MAC) divides nodes into active, passive, and cluster-head roles. The energy consumption is expressed as:
\begin{equation}
E_{BMA} = (E_{CAP}^{BMA} + E_{CFP}^{BMA}) \times l,
\end{equation}
where
\begin{align}
E_{CAP}^{BMA} &= (m+n_w) P_r T_c + (N - m - n_w - 1) P_i T_c + P_t T_{ch}, \\
E_{CFP}^{BMA} &= (n_w + m) P_r T_d + n_w P_t T_d.
\end{align}

Here, $n_w$ represents the number of nodes estimated to transmit during a window with event probabilities (generally higher due to initially fixed static value during simulation runs), while $T_{ch}$ denotes the cluster-head control packet duration. This model captures dynamic adjustments in active-node selection for energy efficiency.

\subsection{EEI-BMA Model (AI-Assisted Variants)}

The proposed Energy-Efficient Intelligent Bit-Map Assisted (EEI-BMA) protocol enhances the traditional BMA by integrating event-probability estimation using an AI predictor. Instead of assuming a fixed event probability for all nodes, EEI-BMA adaptively determines the number of active nodes in each frame based on the predicted event generation probability obtained from a trained neural network.

\subsubsection{1) Predicted Event Probability Mapping}

Let the predicted event probability for node $i$ in the current frame be denoted as $P_{i}^{pred}$.  
The expected number of event-generating nodes in a cluster of size $N$ (excluding continuous monitoring nodes $m$) is given by:
\begin{equation}
n_p = (N - m - 1) \, \bar{P}_{pred},
\label{eq:npred}
\end{equation}
where $\bar{P}_{pred}$ is the mean predicted probability across all nodes.

This term dynamically controls the number of active transmitters per frame, reducing redundant channel access when the predicted event probability is low, and scaling up node participation during bursty events.

\subsubsection{2) Contention Access Period (CAP) Energy}

During CAP, the cluster head transmits the control packet while member nodes either receive, transmit, or stay idle depending on the bitmap activation pattern derived from $P_{i}^{pred}$.  
The total energy consumed during CAP for EEI-BMA is expressed as:
\begin{equation}
E_{CAP}^{EEI} = (m + n_p) P_r T_c + (N - m - n_p - 1) P_i T_c + P_t T_{ch},
\label{eq:cap}
\end{equation}
where:
\begin{itemize}
    \item $P_t$, $P_r$, and $P_i$ denote transmit, receive, and idle power,
    \item $T_c$ is the control packet transmission time,
    \item $T_{ch}$ represents the cluster head control packet duration.
\end{itemize}

The first term quantifies the energy spent by continuous-monitoring and predicted event nodes in reception, the second term accounts for idle listening nodes, and the last term denotes the cluster-head broadcast overhead.

\subsubsection{3) Contention-Free Period (CFP) Energy}

Once the CAP phase concludes, the nodes that were predicted to have an event (based on $n_p$) are allocated dedicated time slots for data transmission.  
The energy consumption during the CFP is:
\begin{equation}
E_{CFP}^{EEI} = n_p P_r T_d + m P_r T_d + n_p P_t T_d,
\label{eq:cfp}
\end{equation}
where $T_d$ is the duration of a single data transmission slot. 
The first two terms capture the reception energy at the cluster head and continuous nodes, while the last term accounts for the transmit energy of active nodes that successfully send their data packets.

\subsubsection{4) Total Energy per Frame}

Combining the above two components, the total energy consumed per frame in EEI-BMA is:
\begin{equation}
E_{frame}^{EEI} = E_{CAP}^{EEI} + E_{CFP}^{EEI}.
\label{eq:frame}
\end{equation}

This value is then scaled by the number of operational frames (or rounds) $l$ to compute the cumulative energy consumption:
\begin{equation}
E_{EEI-BMA} = l \times (E_{CAP}^{EEI} + E_{CFP}^{EEI}).
\label{eq:total}
\end{equation}

\subsubsection{5) AI-Assisted Variants of EEI-BMA}

To evaluate the effect of AI prediction accuracy, three variants of the EEI-BMA model are considered:

\paragraph{(a) True Prediction Model}
In this case, the predicted probability perfectly matches the empirical ground truth obtained from historical event data:
\begin{equation}
n_t = (N - m - 1) \, \bar{P}_{true},
\label{eq:nt}
\end{equation}
and the total energy is:
\begin{equation}
E_{EEI-BMA}^{(true)} = l \times (E_{CAP}^{EEI}(n_t) + E_{CFP}^{EEI}(n_t)).
\end{equation}
This scenario represents an ideal predictor capable of capturing both spatial and temporal correlations precisely.

\paragraph{(b) Best Prediction Model}
When the predictor underestimates the event generation probability (i.e., $\bar{P}_{pred}^{min} < \bar{P}_{true}$), fewer nodes are scheduled for data transmission, possibly leading to retransmissions and reduced energy efficiency:
\begin{equation}
n_b = (N - m - 1) \, \bar{P}_{pred}^{min},
\end{equation}
and
\begin{equation}
E_{EEI-BMA}^{(poor)} = l \times (E_{CAP}^{EEI}(n_b) + E_{CFP}^{EEI}(n_b)).
\end{equation}

\paragraph{(c) Poor Prediction Model}
When the predictor overestimates upcoming events or the environment induces burst synchronization, the network pre-activates a larger number of nodes:
\begin{equation}
n_w = (N - m - 1) \, \bar{P}_{pred}^{max},
\end{equation}
yielding
\begin{equation}
E_{EEI-BMA}^{(best)} = l \times (E_{CAP}^{EEI}(n_w) + E_{CFP}^{EEI}(n_w)).
\end{equation}
Although this case may slightly increase instantaneous energy, it ensures higher readiness and packet delivery during correlated event bursts.

\subsubsection{6) Comparative Performance Index}

To quantify the relative improvement achieved through intelligent prediction, a normalized energy-efficiency ratio is defined as:
\begin{equation}
\eta_{EEI} = \frac{E_{BMA}}{E_{EEI-BMA}},
\label{eq:eta}
\end{equation}
where $\eta_{EEI} > 1$ implies better energy performance compared to the conventional BMA. This ratio emphasizes how predictive adaptation, even under imperfect estimation, contributes to lower overall energy consumption while maintaining comparable throughput and latency.

Equations~(\ref{eq:npred})–(\ref{eq:eta}) collectively describe the adaptive mechanism of EEI-BMA. The dynamic computation of $n_p$ links AI inference directly with MAC-layer scheduling, making the model inherently self-optimizing. The balance between prediction accuracy and transmission scheduling determines the trade-off between energy saving and delivery reliability. Consequently, the EEI-BMA model offers a generalized analytical framework to assess how predictive intelligence impacts network-level energy dynamics. The comparative performance of all MAC schemes is evaluated using:
\begin{equation}
E_{comp} = \{E_{TDMA}, E_{EATDMA}, E_{BMA}, E_{EBMA}^{(poor)}, E_{EBMA}^{(true)}, E_{EBMA}^{(best)}\},
\end{equation}

Each protocol’s energy consumption curve is plotted as a function of various parameters/scenarios, illustrating the trade-off between adaptive scheduling and prediction accuracy. The AI-based EEI-BMA demonstrates superior energy performance under realistic event dynamics due to its ability to anticipate event generation patterns and minimize unnecessary node activation.

\section{Result Analysis and Discussion}
To analyze the performance of the proposed method and comparison with the existing method, the MATLAB-based performance evaluation of the proposed model is done using the CC2420 IEEE 802.15.4 transceiver parameters. The simulation environment considers a clustered WSN/IoT deployment where nodes periodically generate event-driven data. The analysis is conducted under varying conditions of event-generation probability (\(p_{\text{event}}\)), number of nodes (\(N\)), number of continuous monitoring nodes (\(m\)), and data packet duration (\(T_d\)). The complete simulation parameters used in the experiments are summarized in Table~\ref{tab:sim_params}.

\begin{table}[!t]
\centering
\footnotesize
\caption{Simulation Parameters Used in MATLAB-Based Evaluation}
\label{tab:sim_params}
\begin{tabular}{|p{6cm}|c|}
\hline
\textbf{Parameter} & \textbf{Value} \\ \hline
Transceiver model & CC2420 (IEEE 802.15.4) \\ \hline
Transmit power \(P_t\) & 52.2 mW \\ \hline
Receive power \(P_r\) & 56.4 mW \\ \hline
Idle power \(P_i\) & 1.42 mW \\ \hline
Energy check power \(P_e\) & 2.0 mW \\ \hline
Control packet duration \(T_c\) & 1.5625 ms \\ \hline
Data packet duration \(T_d\) & 62.5 ms (varied) \\ \hline
Cluster-head control time \(T_{ch}\) & 1.5625 ms \\ \hline
Total frames \(l\) & 20 \\ \hline
Number of nodes (\(N\)) & 20–50 (varied) \\ \hline
Continuous monitoring nodes (\(m\)) & 1–9 \\ \hline
Event-generation probability (\(p_{\text{event}}\)) & 0.1–0.5 (varied) \\ \hline
Neural predictor hidden units & 64 \\ \hline
Sequence length & 8 frames \\ \hline
\end{tabular}
\end{table}

\subsection{Probability Estimation Analysis}

In order to validate the accuracy of the proposed event–generation probability estimator, a dedicated experiment was carried out using the synthetic dataset produced from the temporal–spatial event model.

\begin{figure*}
    \centering
    \includegraphics[width=0.95\linewidth]{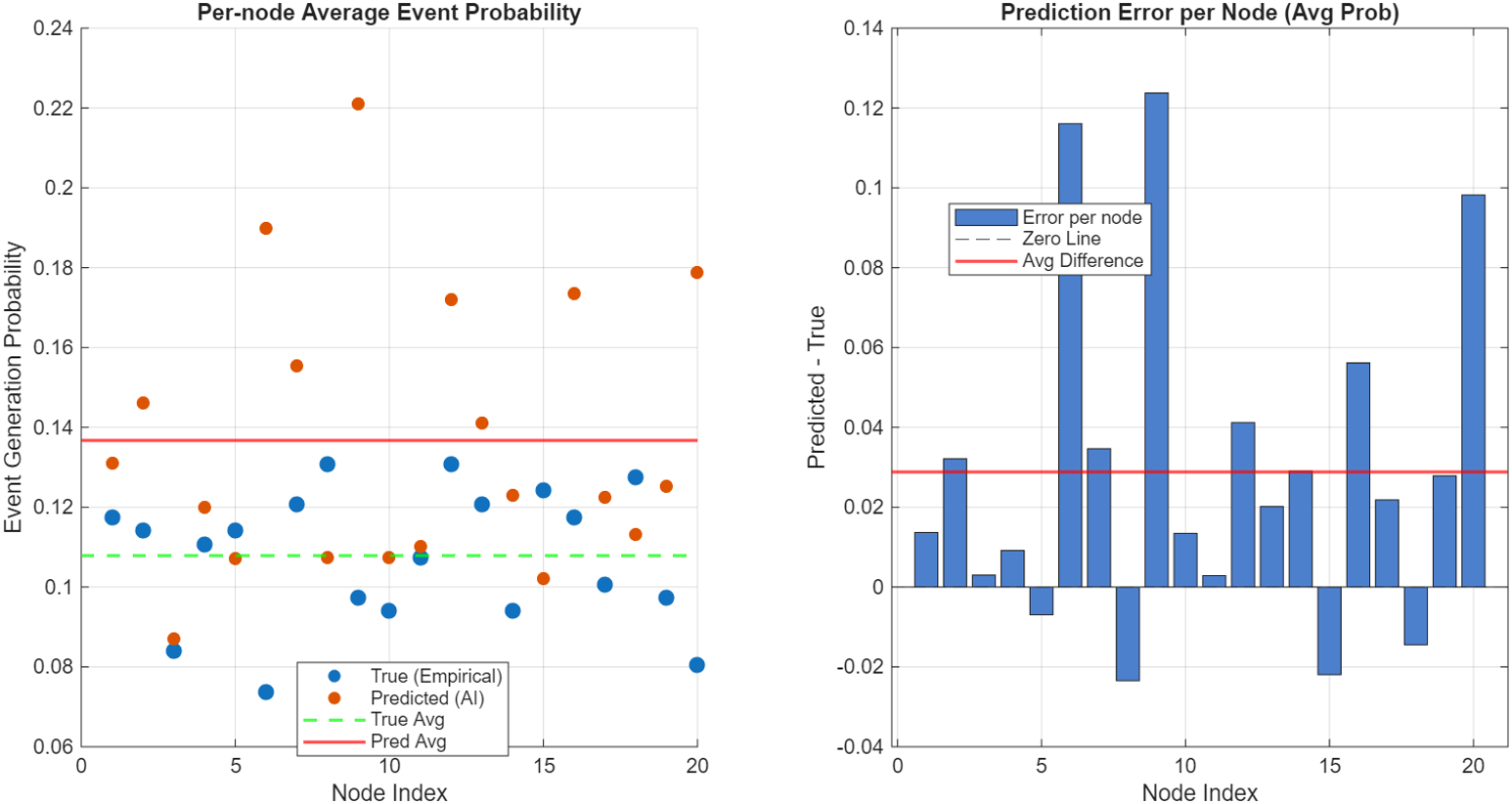}
    \caption{Per-node probability estimation results: (left) true vs.\ predicted event-generation probability; (right) node-wise estimation error.}
    \label{fig:prob_estimation}
\end{figure*}

The neural predictor was trained on a sequence length of eight frames and evaluated on the remaining test frames. Figure~\ref{fig:prob_estimation} presents two complementary visualizations: (i) the per–node average event-generation probabilities and (ii) the corresponding prediction error for each node.

The left subplot in Fig.~\ref{fig:prob_estimation} compares the empirical (true) event-generation probability of each node with the probability predicted by the AI-based estimator. The dashed green line represents the global empirical average across all nodes, while the solid red line denotes the global predicted average. The right subplot shows the node-wise prediction error computed as the difference between predicted and true probability values. A positive value indicates overestimation, whereas a negative value indicates underestimation.

\subsection{Impact of Event-Generation Probability (\(p_{\text{event}}\))}
Figure~\ref{fig:p} presents the energy consumption of various MAC protocols under different event-generation probabilities (\(p\)), demonstrating the efficiency of the proposed EEI-BMA framework. Traditional-TDMA incurs the highest and nearly constant energy cost due to fixed slot allocation, while Energy-Aware TDMA reduces the consumption slightly but still scales inefficiently as traffic increases. Traditional-BMA performs well at low \(p\) but its energy usage rises sharply at higher loads because of contention and retransmissions. In contrast, all three variants of EEI-BMA (poor, true, and best prediction) consistently consume less energy across all values of \(p\), enabled by prediction-guided duty cycling and selective activation. Even with inaccurate probability estimates, EEI-BMA outperforms both TDMA and Traditional-BMA, whereas true and best prediction achieve the lowest consumption, offering up to 40--50\% savings over TDMA and 30--40\% over Energy-Aware TDMA, particularly when \(p\) is low. These results validate that integrating probabilistic prediction into MAC scheduling significantly enhances energy efficiency while maintaining robustness under imperfect estimation.

\begin{figure*}
    \centering
    \includegraphics[width=0.95\linewidth]{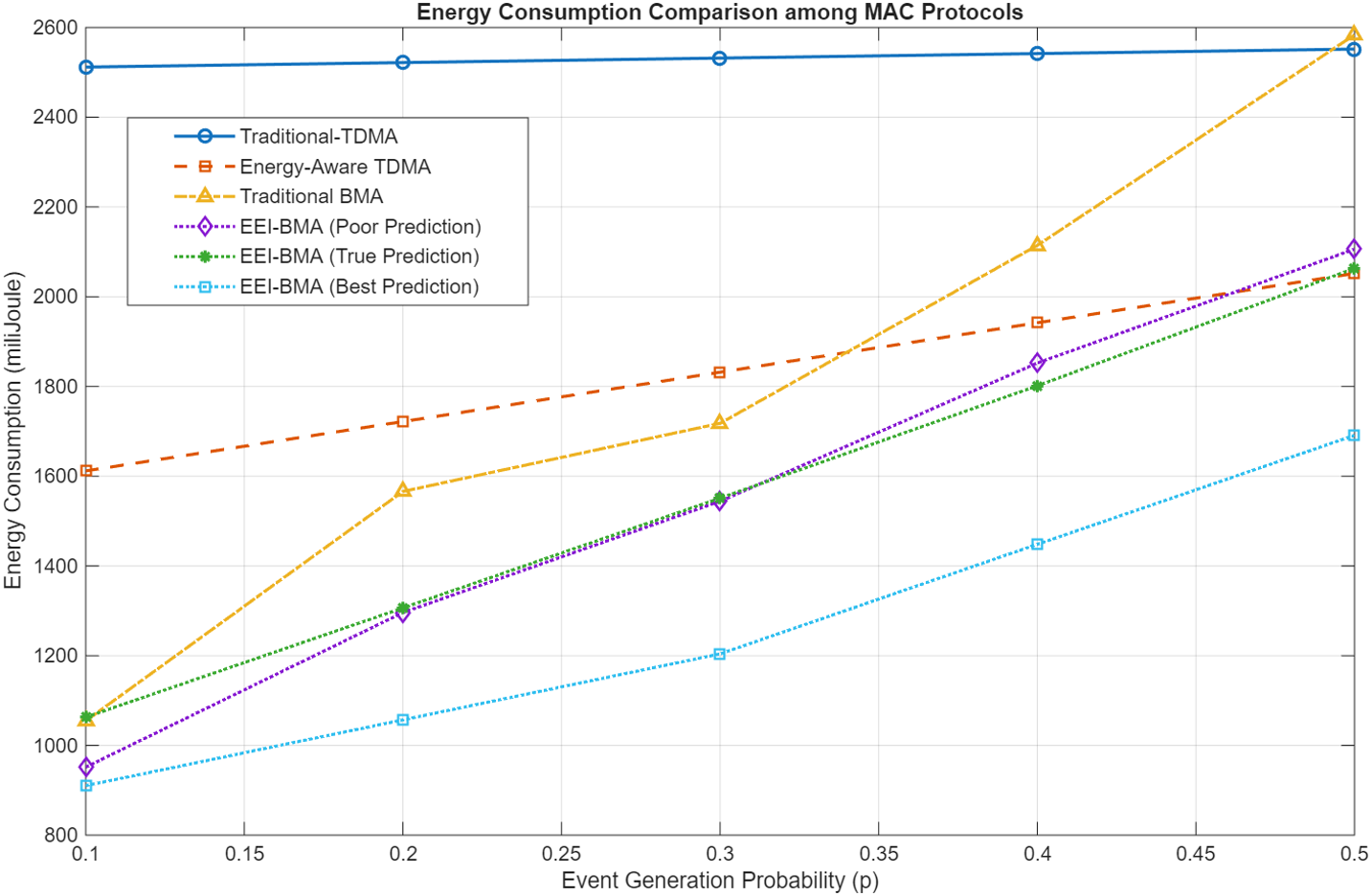}
    \caption{Energy consumption comparison of MAC protocols under varying event-generation probabilities}
    \label{fig:p}
\end{figure*}

\subsection{Impact of Number of Nodes (\(N\))}
Figure~\ref{fig:N} illustrates the variation in energy consumption with respect to the number of nodes (\(N\)) for different MAC protocols. As expected, Traditional-TDMA shows the highest and almost linear increase in energy usage because each node is assigned a dedicated slot regardless of traffic, resulting in large idle-listening overhead as the network scales. Energy-Aware TDMA performs better but still increases significantly due to its periodic wake-up structure and limited adaptability. Traditional-BMA initially offers lower consumption but becomes less efficient as \(N\) grows due to increasing contention and retransmissions. In contrast, all variants of the proposed EEI-BMA framework demonstrate substantially lower and more controlled energy growth. The Best Prediction case achieves the lowest consumption across all \(N\), while even Poor Prediction remains superior to TDMA and competitive with BMA. The True Prediction scenario shows consistent mid-range performance closely aligned with analytical expectations. The zoomed inset further highlights the small performance gap between EEI-BMA variants at higher node densities, confirming that the protocol remains robust and scalable even under imperfect probability estimation. 

\begin{figure*}
    \centering
    \includegraphics[width=0.95\linewidth]{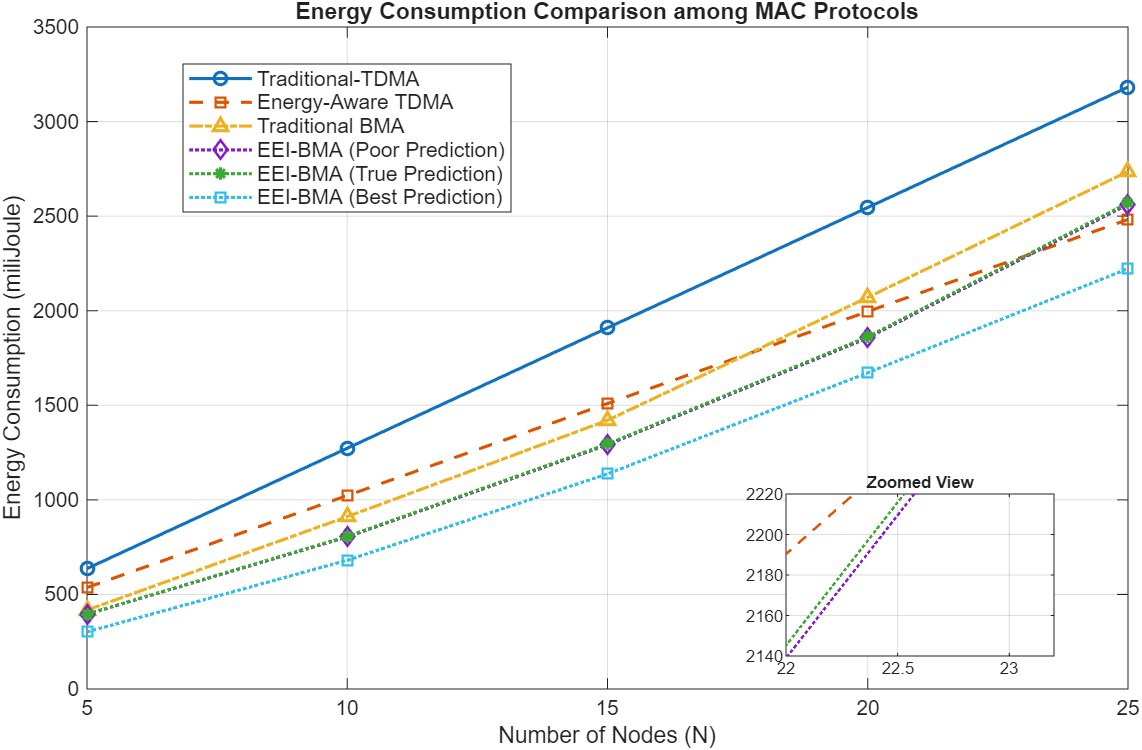}
    \caption{Energy consumption of MAC protocols for varying number of nodes}
    \label{fig:N}
\end{figure*}

\subsection{Impact of Continuous Monitoring Nodes (\(m\))}
Figure~\ref{fig:m} illustrates how energy consumption varies with the number of continuous monitoring nodes (\(m\)) for different MAC protocols. As expected, Traditional-TDMA exhibits the highest and almost flat energy consumption because all nodes remain active in every frame regardless of their monitoring role. Energy-Aware TDMA reduces energy usage but still scales inefficiently with larger \(m\) due to periodic active-duty schedules. Traditional-BMA initially shows competitive performance but becomes less efficient with higher monitoring intensity, as contention and retransmissions increase. In contrast, all EEI-BMA variants consistently achieve lower energy consumption across the range of \(m\), demonstrating the benefit of prediction-driven selective activation. The Best Prediction case shows the lowest consumption due to optimal scheduling, while the True and Poor Prediction cases remain well below the baselines. The zoomed inset highlights that even for small values of \(m\), the difference between EEI-BMA variants is minimal, indicating strong robustness to prediction inaccuracies.

\begin{figure*}
    \centering
    \includegraphics[width=0.95\linewidth]{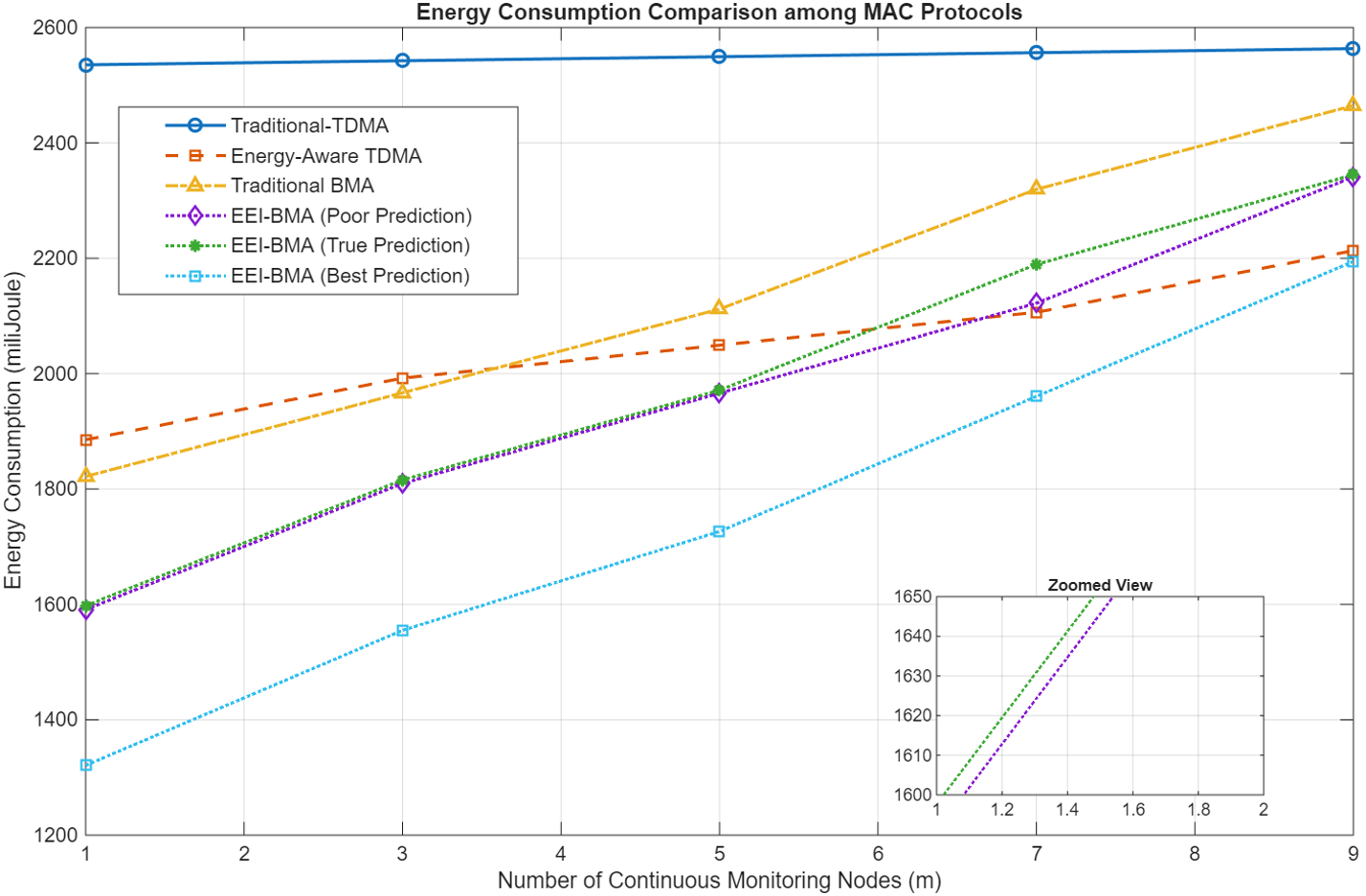}
    \caption{Energy consumption of MAC protocols under varying numbers of continuous monitoring nodes}

    \label{fig:m}
\end{figure*}

\subsection{Impact of Packet Duration (\(T_d\))}
Figure~\ref{fig:Td} shows the impact of packet size (\(T_d\)) on the energy consumption of different MAC protocols. As expected, Traditional-TDMA exhibits the highest and steepest growth in energy usage because every increase in packet duration directly extends the active radio time for all nodes. Energy-Aware TDMA performs moderately better but still scales inefficiently due to its fixed periodic transmission structure. Traditional-BMA initially performs closer to the energy-aware scheme but becomes increasingly inefficient with larger packets because longer contention periods lead to more collisions and retransmissions. In contrast, all variants of the proposed EEI-BMA protocol demonstrate significantly lower and more controlled energy growth across all \(T_d\) values. The Best Prediction case achieves the minimum energy consumption due to optimal sleep–wake scheduling, while the True and Poor Prediction cases remain consistently below all baseline protocols. The zoomed inset further highlights that even with imperfect prediction, the energy gap between EEI-BMA variants remains small, confirming robustness to estimation errors. 

\begin{figure*}
    \centering
    \includegraphics[width=0.95\linewidth]{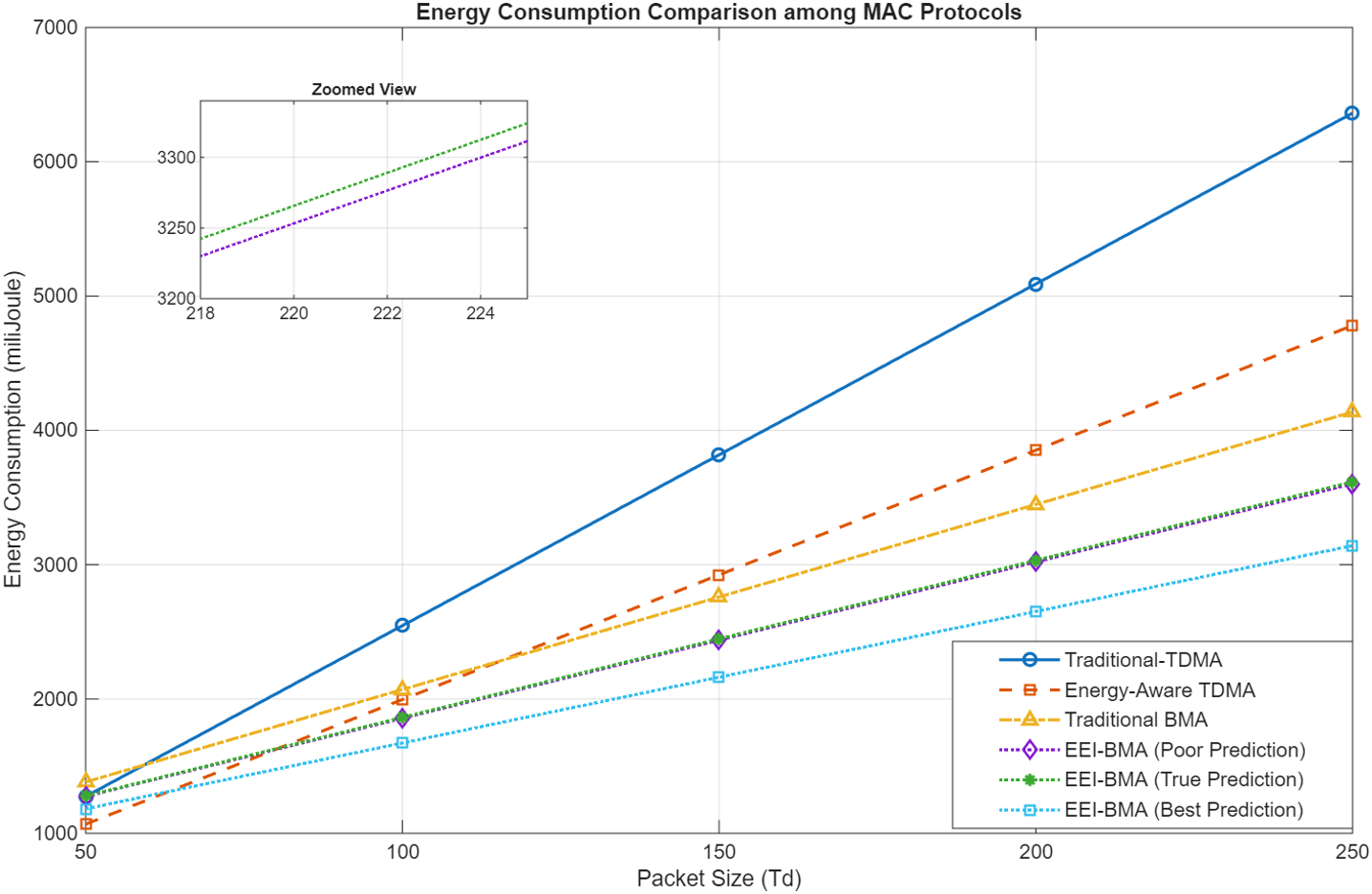}
    \caption{Energy consumption of MAC protocols under varying packet sizes}
    \label{fig:Td}
\end{figure*}

\subsection{Comparative Analysis of Protocols}

Across all evaluated scenarios—including variations in event-generation probability ($p$), number of nodes ($N$), packet size ($T_d$), and the number of continuous monitoring nodes ($m$)—the proposed EEI\text{-}BMA protocol demonstrates consistent and significant reductions in energy consumption compared with existing MAC schemes. Numerical observations show that EEI\text{-}BMA (Best Prediction) achieves approximately 35--45\% lower energy usage than Traditional--TDMA, and 22--30\% savings over Energy-Aware TDMA under typical sensing loads. Even under imperfect prediction conditions, the EEI\text{-}BMA (Poor Prediction) model maintains 18--28\% lower consumption than Traditional--BMA and 15--22\% lower consumption than Energy-Aware TDMA. For higher node densities (e.g., $N=25$), EEI\text{-}BMA reduces energy consumption by nearly 40\% relative to TDMA, while experiments with larger packet sizes ($T_d=250$~bytes) show a sustained 30--38\% improvement. Similarly, with increasing continuous monitoring load, EEI\text{-}BMA delivers 27-35\% energy savings over TDMA and 12-20\% over BMA. Overall, the close performance of the Best and True prediction models, along with the consistently superior performance of the Poor prediction model over all classical baselines, confirms the robustness, scalability, and efficiency of the proposed EEI\text{-}BMA framework across diverse operating conditions.

\section{Conclusion}\label{sec13}

In the proposed work an EEI-BMA protocol is presented. The proposed EEI-BMA an energy-efficient and prediction-assisted MAC framework designed for heterogeneous IoT and WSN environments. The MATLAB-based simulation is conducted that proves the proposed scheme achieves substantial energy savings across diverse operating conditions. In particular, EEI-BMA (Best Prediction) reduced energy consumption by approximately 35--45\% compared with Traditional-TDMA, 22--30\% relative to Energy-Aware TDMA, and 18-28\% compared with Traditional-BMA, while even the Poor Prediction variant consistently outperformed all baseline protocols. These improvements validate the effectiveness of incorporating event-probability forecasting into MAC-layer scheduling. Future research may extend this work by integrating reinforcement learning for adaptive probability estimation, exploring cross-layer optimization with routing and clustering mechanisms, evaluating performance under mobility and interference-rich environments, and deploying the protocol on real testbeds or hardware-in-the-loop platforms to further validate scalability and robustness in practical large-scale IoT systems.

\newpage

\section*{List of Abbreviations}
\noindent Abbreviations of notation are given in Table \ref{tab:abbreviations}.

\begin{table}[h]
\label{tab:abbreviations}
\caption{List of Abbreviations}
\label{tab:abbreviations}
\begin{tabular}{ll}
\hline
\textbf{Abbreviation} & \textbf{Description} \\ \hline
IoT & Internet of Things \\ \hline
M2M & Machine-to-Machine Communication \\ \hline
MAC & Medium Access Control \\ \hline
ML & Machine Learning \\ \hline
WSN & Wireless Sensor Network \\ \hline
URLLC & Ultra-Reliable Low-Latency Communication \\ \hline
TDMA & Time Division Multiple Access \\ \hline
CSMA & Carrier Sense Multiple Access \\ \hline
EA-TDMA & Energy-Aware Time Division Multiple Access \\ \hline
BMA & Bit-Map Assisted MAC \\ \hline
EEI-BMA & Energy-Efficient Intelligent Bit-Map Assisted MAC \\ \hline
CAP & Contention Access Period \\ \hline
CFP & Contention-Free Period \\ \hline
NN & Neural Network \\ \hline
BCE & Binary Cross-Entropy \\ \hline
RMSE & Root Mean Square Error \\ \hline
SAGIN & Space-Air-Ground Integrated Network \\ \hline
MTCD & Machine-Type Communication Device \\ \hline
FL & Federated Learning \\ \hline
IDS & Intrusion Detection System \\ \hline
SDN & Software Defined Network \\ \hline
QoS & Quality of Service \\ \hline
IIoT & Industrial Internet of Things \\ \hline
E2EC & End-to-End Confidentiality \\ \hline
HE & Homomorphic Encryption \\ \hline
AC & Access Control \\ \hline
CR & Collusion Resistance \\ \hline
CL & Clustering \\ \hline
EE & Energy Efficiency \\ \hline
PP & Predictive Power \\ \hline
AR & Attack Resilience \\ \hline
\end{tabular}
\end{table}

\section*{Declarations}

\subsection*{Data Availability}
There is no data available.

\subsection*{Funding}
This research is not funded by any agency. 

\subsection*{Conflict of interest/Competing interests}
The authors have no competing interests to declare that are relevant to the content of this article. This research did not receive any specific grant from funding agencies in the public, commercial, or not-for-profit sectors.

\subsection*{Author Contribution}

Jesmine Damilola Omonori conceived the study, designed the methodology, conducted the analysis, and interpreted the results. Iyanu Tomiwa Durotola and Godspower Paul Osilama drafted the manuscript, revised it critically for important intellectual content, and approved the final version for publication.

\subsection*{Acknowledgements}
Not applicable.

\subsection*{Authors’ information}
Not applicable.


\bibliography{sn-bibliography}

\vspace{3cm}

\noindent \textbf{Figures, Figure Title and Legend Section}
\addtocounter{figure}{-5} 
\begin{figure*}[h]
    \centering
    \includegraphics[width=0.95\linewidth]{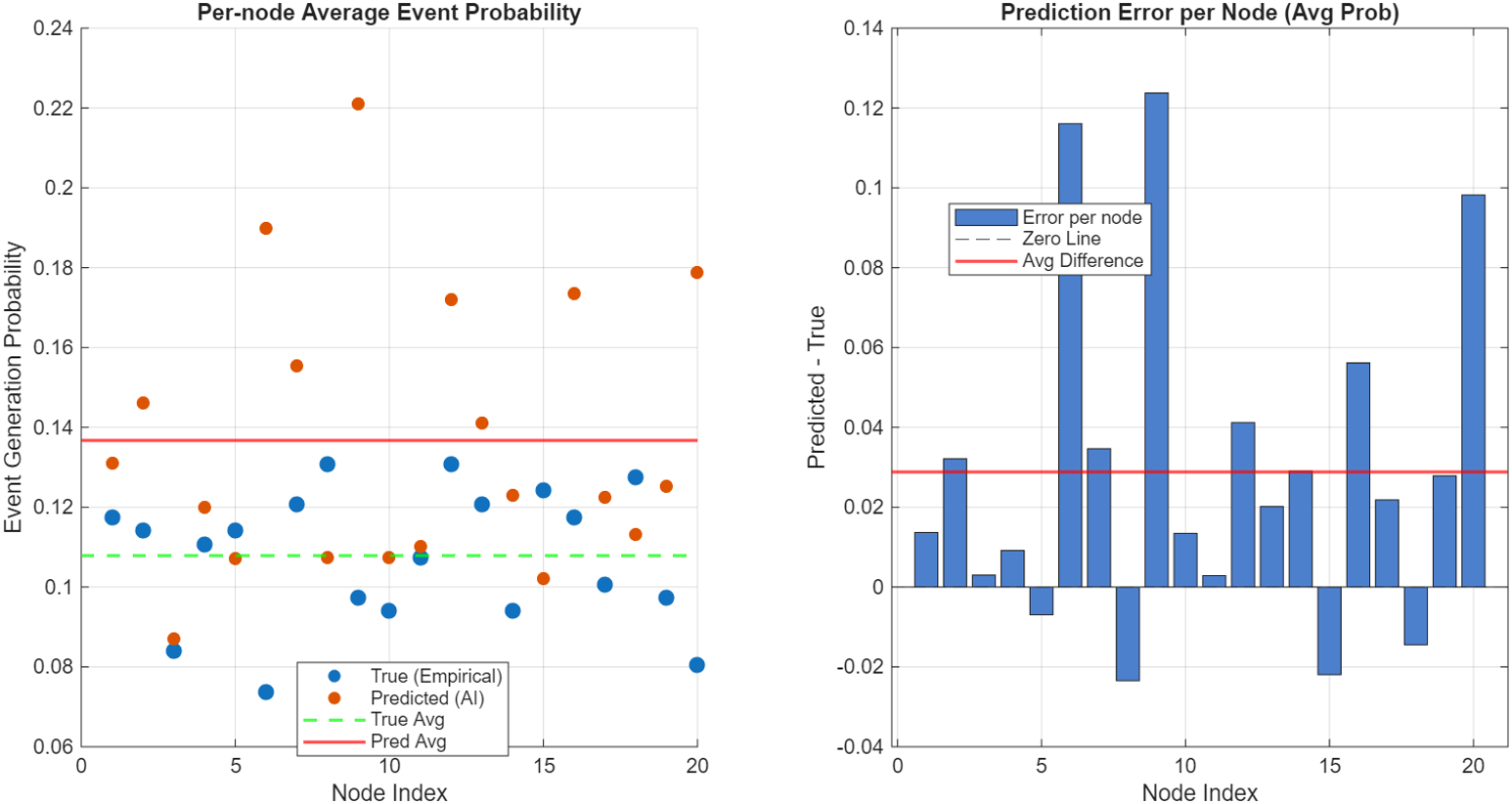}
    \caption{Per-node probability estimation results: (left) true vs.\ predicted event-generation probability; (right) node-wise estimation error.}
    \label{fig:prob_estimation}
\end{figure*}

\noindent \textbf{Legend:} Fig. \ref{fig:prob_estimation} represents Blue – True probability; Red – Predicted probability; Green dashed – Global empirical average; Red solid – Global predicted average; Right panel – Node-wise prediction error (positive: overestimation, negative: underestimation).

\begin{figure*}[h]
    \centering
    \includegraphics[width=0.7\linewidth]{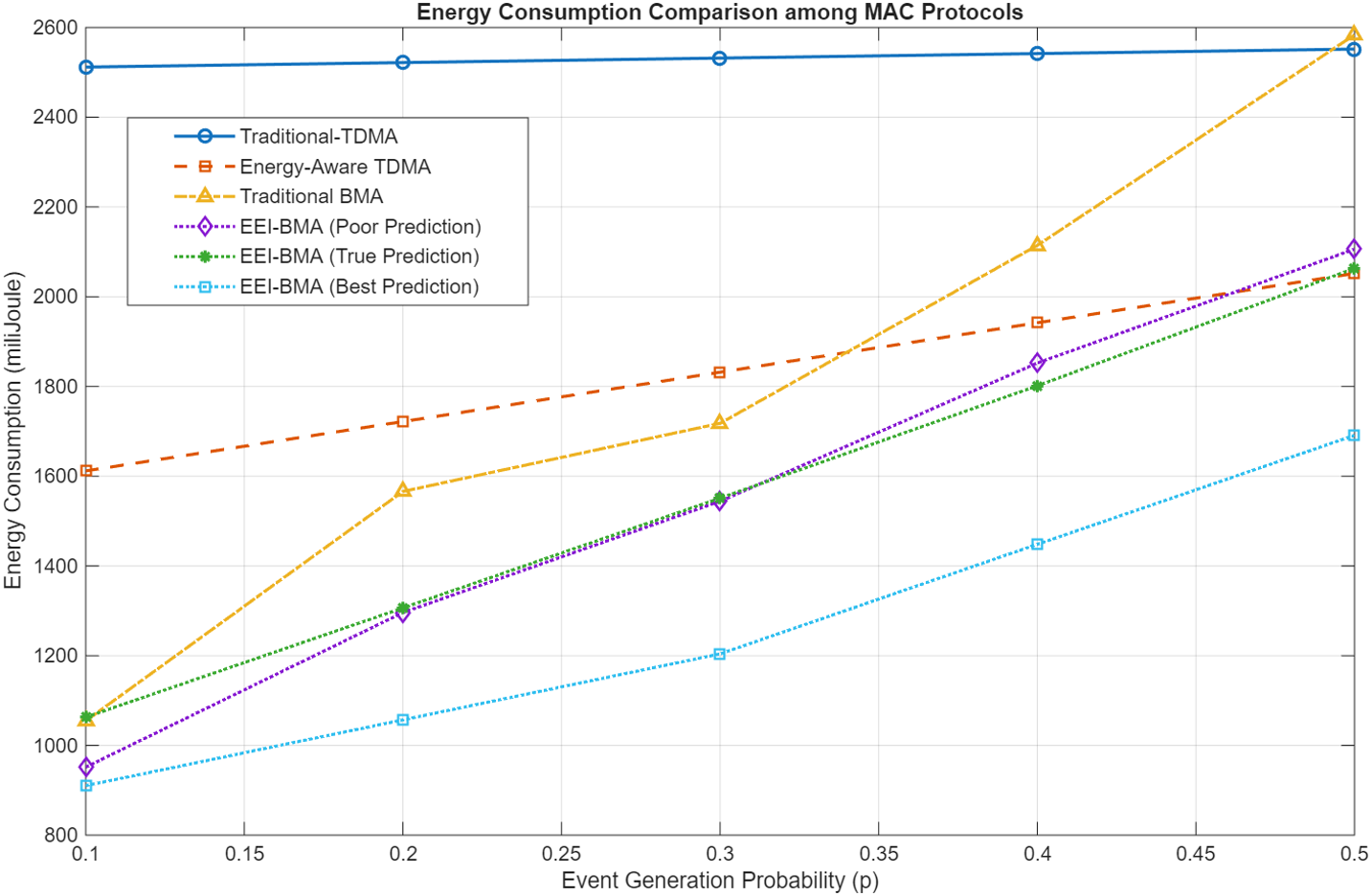}
    \caption{Energy consumption comparison of MAC protocols under varying event-generation probabilities}
    \label{fig:p}
\end{figure*}

\noindent \textbf{Legend:} Fig. \ref{fig:p}  includes six MAC protocol variants: conventional TDMA with fixed slot allocation; EA-TDMA, which incorporates buffer-state checking for improved energy efficiency; traditional BMA with bitmap-based dynamic activation; and three AI-assisted EEI-BMA variants. EEI-BMA (Poor) represents underestimated event probability, EEI-BMA (True) uses empirical ground-truth probability, and EEI-BMA (Best) represents proactive or higher predicted probability under burst-aware conditions.

\begin{figure}
    \centering
    \includegraphics[width=0.7\linewidth]{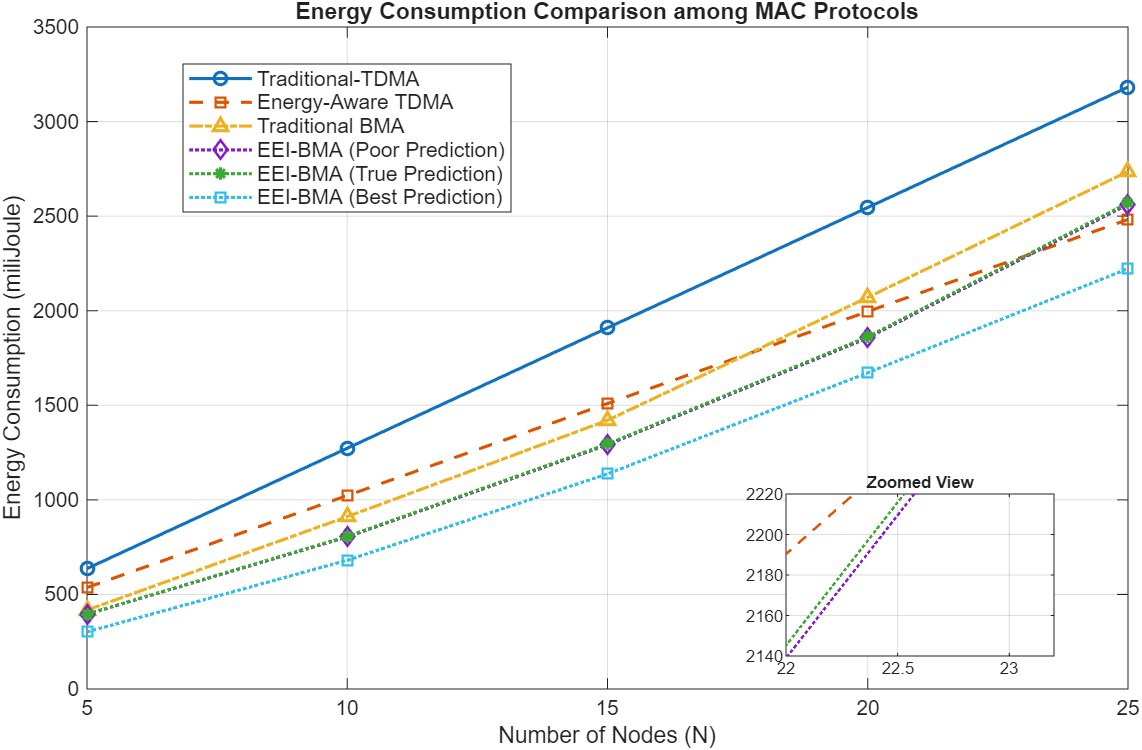}
    \caption{Energy consumption of MAC protocols for varying number of nodes}
    \label{fig:N}
\end{figure}

\noindent \textbf{Legend:} Fig. \ref{fig:N} includes TDMA with static scheduling, EA-TDMA with adaptive energy-aware slot management, and traditional BMA using bitmap-based coordination. The proposed EEI-BMA is evaluated under three prediction scenarios: Poor (conservative/underestimated activation), True (accurate empirical prediction), and Best (proactive or higher predicted activation), demonstrating the impact of intelligent scheduling as the number of nodes increases.

\begin{figure}
    \centering
    \includegraphics[width=0.7\linewidth]{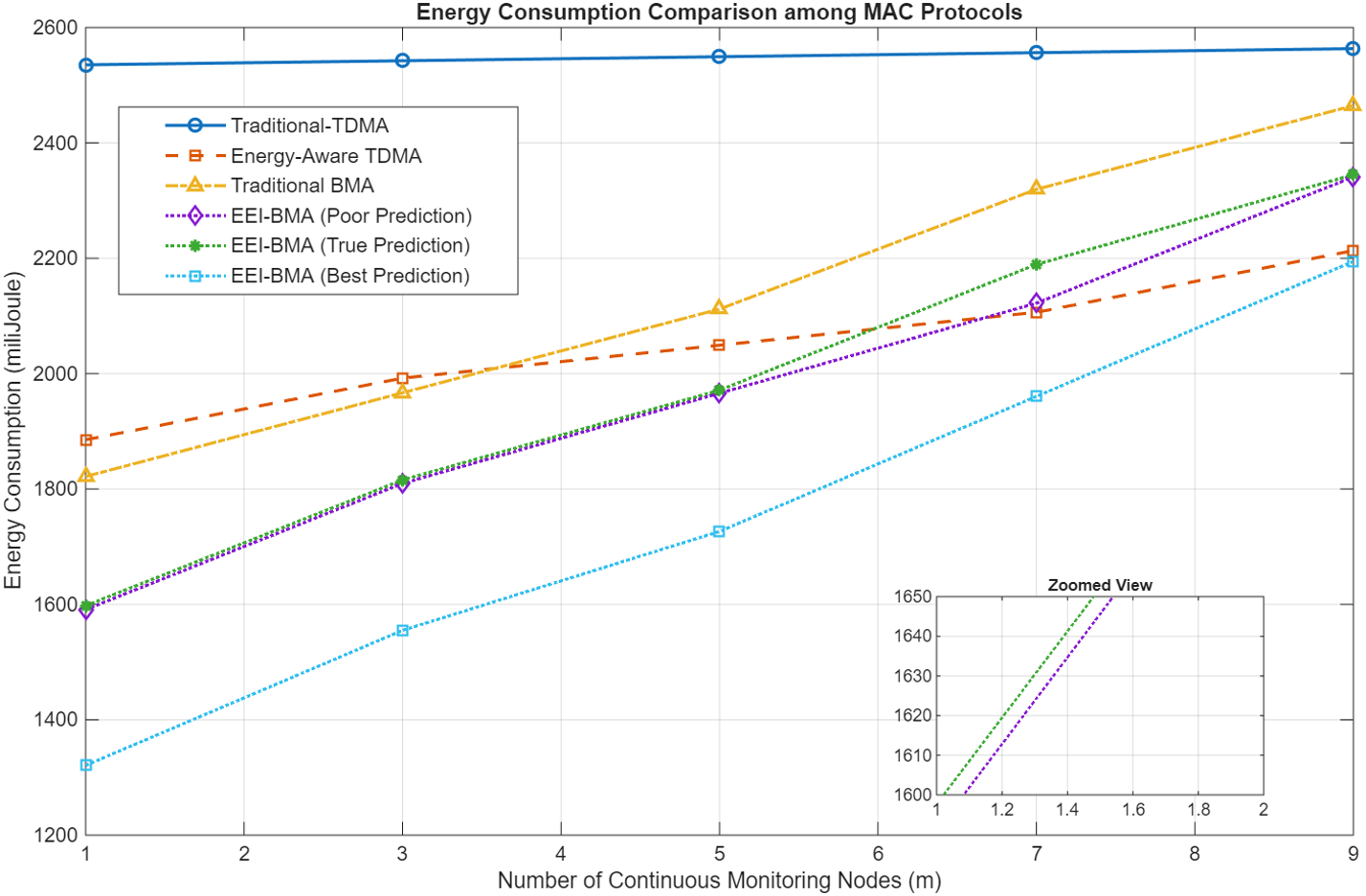}
    \caption{Energy consumption of MAC protocols under varying numbers of continuous monitoring nodes}
    \label{fig:m}
\end{figure}

\noindent \textbf{Legend:} Fig. \ref{fig:m} represents TDMA with uniform slot allocation, EA-TDMA with energy-check mechanisms, and traditional BMA with fixed probability-based activation. The EEI-BMA variants include Poor prediction (lower estimated event probability), True prediction (empirical probability), and Best prediction (higher predicted activity), highlighting how adaptive AI-driven scheduling responds to varying numbers of continuous monitoring nodes.

\begin{figure*}[h]
    \centering
    \includegraphics[width=0.7\linewidth]{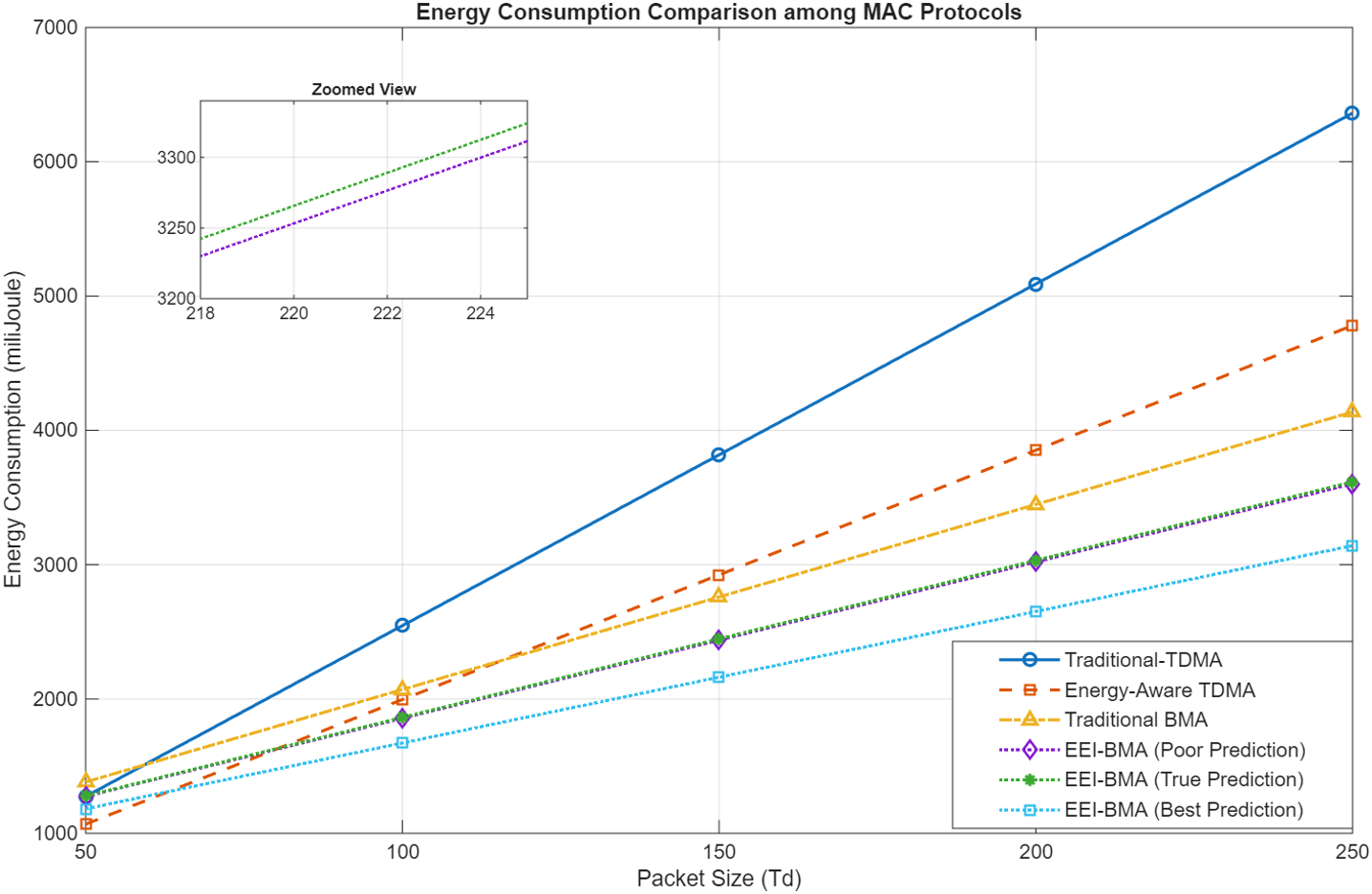}
    \caption{Energy consumption of MAC protocols under varying packet sizes}
    \label{fig:Td}
\end{figure*}

\noindent \textbf{Legend:} Fig. \ref{fig:Td} curves correspond to TDMA, EA-TDMA, and traditional BMA, along with three EEI-BMA prediction cases. The Poor case reflects underestimated traffic conditions, the True case reflects accurate probability estimation, and the Best case reflects proactive prediction under higher expected event activity. These variants illustrate how packet duration influences energy consumption across static and AI-assisted MAC schemes.

\end{document}